# Optical and thermal analysis of the light-heat conversion process employing an antenna-based hybrid plasmonic waveguide for HAMR

Nicolás Abadía,[1,2,3,5,*] Frank Bello,[1,2,3] Chuan Zhong,[1,2,3] Patrick Flanigan,[1,2,3] David M. McCloskey,[1,2,3] Christopher Wolf,[4] Alexander Krichevsky,[4] Daniel Wolf,[4] Fenghua Zong,[4] Alireza Samani,[5] David V. Plant,[5] and John F. Donegan[1,2,3]

[1]*Photonics Group, School of Physics, Trinity College, Dublin 2, Ireland*
[2]*Centre for Research on Adaptive Nanostructures and Nanodevices (CRANN), Trinity College Dublin, Dublin 2, Ireland*
[3]*AMBER Research Centre, Trinity College Dublin, Dublin 2, Ireland*
[4]*Western Digital Corporation (WDC), San Jose, CA, USA*
[5]*Department of Electrical and Computer Engineering, McGill University, Montreal, QC, H3A 2A7, Canada*
*\*nicolas.abadiacalvo@mcgill.ca*

**Abstract:** We investigate a tapered, hybrid plasmonic waveguide which has previously been proposed as an optically efficient near-field transducer (NFT), or component thereof, in several devices which aim to exploit nanofocused light. We numerically analyze how light is transported through the waveguide and ultimately focused via effective-mode coupling and taper optimization. Crucial dimensional parameters in this optimization process are identified that are not only necessary to achieve maximum optical throughput, but also optimum thermal performance with specific application towards heat-assisted magnetic recording (HAMR). It is shown that existing devices constructed on similar waveguides may benefit from a heat spreader to avoid deformation of the plasmonic element which we achieve with no cost to the optical efficiency. For HAMR, our design is able to surpass many industry requirements in regard to both optical and thermal efficiency using pertinent figure of merits like 8.5% optical efficiency.

## References and links

# 1. Introduction

Heat-assisted magnetic recording, or HAMR, has been proposed on the Advanced Storage Technology Consortium's (ASTC) roadmap as the next-generation recording technique to continue scaling hard disk drive (HDD) magnetic storage [1]. This technology is projected to increase storage capabilities roughly 10-fold, and prototype devices have demonstrated areal bit densities as large as 2.6 Tb/in$^2$ [1–3]. HAMR utilises high coercivity materials such as FePt to allow writing of stable bits at higher densities. To effectively write in these materials, an area of the recording media smaller than $50 \times 50$ nm$^2$ must be heated above the Curie temperature of the recording layer ($\approx$750 K for bulk L1$_0$ FePt) [4]. Above this temperature, the coercivity decreases sufficiently to allow the magnetization within the layer to be controlled by a perpendicular magnetic recording (PMR) write head. After writing, the area rapidly cools 'freezing' the applied magnetization and forming a stored bit of information. This process must proceed while keeping the temperature of the read/write head and adjacent bits as close as possible to ambient temperatures [5].

There are multiple routes to generate localised heating, including microwave and scanning probe heating, however one of the most promising techniques uses optical excitation. In this case the output of a near infrared (NIR) diode laser is coupled to an integrated dielectric waveguide which delivers the optical power to a plasmonic element known as a near-field transducer (NFT). The NFT then focuses the plasmonic mode to the air-bearing surface (ABS), i.e. fine tip of the taper, and lastly couples non-radiatively to the lossy metallic recording media generating a localised, sub-diffraction heat source. We aim to improve the optimization process and identify key parameters which propagate light from *(1)* dielectric core to plasmonic NFT and then *(2)* plasmonic NFT to recording media where light is efficiently converted to heat energy. A range of promising NFT designs have been proposed and prototyped including the aperture-based designs such as the C-, E-, and H-shaped apertures [6–8], bow-tie apertures [9, 10], and antenna-based designs such as triangular and lollipop transducers [11–15] to name a few. There are basically two ways of exciting the NFT in these kinds of structures: evanescent-coupling from dielectric waveguides [16, 17] or direct end-fire coupling [18–20]. This article specifically examines the functionality and performance when evanescent coupling a dielectric waveguide to a metallic film tapered in one dimension (see Fig. 1) similar to a number of antennas proposed for HAMR and commonly used in nanoheating and nanofocusing devices [21].

A key design parameter for HAMR is the optical absorption efficiency, which we define as the power dissipated as heat in the recording layer covering an area of $50 \times 50$ nm$^2$ divided by the optical power incident on the dielectric waveguide. Typical optical efficiencies are in the range of 1-8% though definitions of efficiency tend to vary widely, e.g. ones that consider absorption within the entire media stack to those that consider the limited area we have specified, and which contain the bulk of the energy utilized in the bit writing process. Using identical definitions, Singh *et al.* [22] model a similar NFT and reported an optical efficiency of 6.4%, whereas we find a maximum optical efficiency of 8.5% using comparable dimensions although at a different wavelength. Also using like definitions, the nanobeak antenna which also incorporates a tapered metallic NFT, but with a refined tip tapered in multiple dimensions [3]. In this case, Matsumoto *et al.* optimized their design using a slightly different set of parameters and ultimately settled on a taper length many times shorter than we have. We discuss reasons for these deviations in Section 3 of this manuscript.

This article also places a high priority on thermal efficiency to ensure transition sharpness between successive bits (~thermal gradients) [4], sufficient areal bit density (~thermal spot size) [23], and extended NFT lifetime (~peak temperatures) with most notably increasing NFT temperature and protrusion being a major impediment to scaling current HAMR designs [24]. In outlining the light-heat conversion process, the extended lifetime of the NFT and areal bit density are given precedence over other parameters used to quantify HAMR performance. Thermal data may not always be reported for NFTs suggested for HAMR, and none have been found reporting on the thermal behavior of the multilayered hybrid design proposed in this investigation; partly chosen for being straightforward to fabricate or make alterations to. We adopt the definition of thermal efficiency defined in [25] which provides a thorough review on many of NFTs mentioned above (triangle, lollipop, E-antennas, bow-tie and C apertures) though no magnetic write pole is included as we have done which can significantly alter results. They supply *(1)* thermal spot size data, *(2)* cross and down track temperature gradients, *(3)* maximum temperatures, along with *(4)* the peak temperature rises per unit of input power (K/mW) in the recording medium compared to the NFT. The latter being critical for extending the lifetime of the NFT. It should be noted that although optical and thermal efficiencies are quoted for comparison with other designs, media and write pole parameters can frequently vary and may be further optimized for additional improvement. Beginning in Section 2, we herein focus on the process taken to efficiently design a hybrid plasmonic waveguide for converting light to heat energy and identify key parameters for optimal HAMR performance. A comparison of similar NFT designs for HAMR is included in Section 3 along with concluding remarks.

## 2. NFT design

Gold (Au) is the most commonly proposed plasmonic material for the NFT, which causes issues for the long-term stability of the devices. Au is soft and has a high surface diffusivity leading to a tendency for the NFT to deform when heated; lowering its energy state at elevated temperatures [26]. Resistive losses in the metallic taper can generate a heat source which considerably raises the steady state NFT temperature and therefore accelerate its degradation. This has led to the investigation of alternative plasmonic materials, such as highly doped Silicon

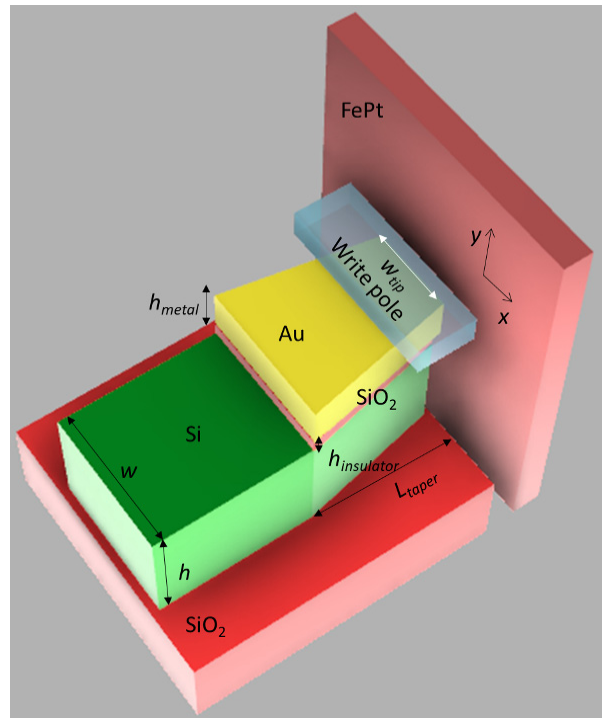

Fig. 1. Initial structure for the NFT, a Si strip waveguide with evanescent coupling to an Au plasmonic structure. The position of the write pole and the FePt disk are also shown. The width and height of the dielectric waveguide are fixed at $w$ = 450nm and $h$ = 250nm respectively. The dimensions of the tapered region must be optimized in the simulations

[27], or refractory materials such as Titanium Nitride [28] and Zirconium Nitride [29]. It seems however that the superior plasmonic performance of Au is still desirable over these alternative materials. Minimizing its temperature rise in the NFT is a key goal to improving the NFT stability and ultimate lifetime of many HAMR devices.

We begin in Sec. 2.1 by numerically investigating the tapered NFT design in Fig. 1 based on evanescent coupling from a Silicon strip waveguide to a relatively short plasmonic taper (dielectric core → NFT optimization). This design is considered a hybrid plasmonic waveguide [30] in which only one side (the bottom side) of the metal supports the plasmon mode and a substantial proportion of the plasmonic mode's energy is kept within the thin insulator spacer layer. There are two advantages of using Silicon as the dielectric core material compared with other dielectric materials such as SiO₂ or Ta₂O₅. Firstly, the high refractive index of Silicon enables better light confinement and coupling to high effective index plasmon modes as will be shown below. Secondly, and more importantly, the thermal conductivity of Si ($\kappa$ = 131 W/m/K) is much larger than Ta₂O₅ ($\kappa$ = 0.45 W/m/K) resulting in a significantly lower steady-state NFT temperature for equal amounts of dissipated power [31, 32]. The modeling we employ uses finite-difference time-domain (FDTD) and finite-element method (FEM) numerical simulations to study the optical and thermal properties, respectively. In Sec. 2.2, we perform calculations of the thermal parameters within the NFT and the recording layer while optimizing the taper dimensions and input power for maximum thermal efficiency (NFT → recording layer optimization). We ultimately propose the addition of a heat spreader to be included with antenna-based NFT designs to reduce the power loss density (heat source) and therefore the steady state operating temperature of the NFT. This heat spreader is designed in such a way so as not to compromise the optical performance of the NFT.

For the optical analysis, a commercial-grade simulator based on the FDTD method and eigenmode solver, Lumerical, was used to simulate the 3D Helmholtz wave equation [33, 34]. Calculations were performed every 0.3 nm in all simulations to include enough data points within the ultra-thin layers of the media. Material parameters used in simulations are shown in Table 1 with waveguide parameters taken from [31, 32] and media layers listed in sequential order. The Air/Lube layer is a mixture of heated air, lubricant materials, and contaminates such as Carbon from the Carbon overcoat (COC) [35, 36]. The structure in Fig. 1 consists of an input strip Si waveguide with the hybrid waveguide/NFT placed at the end facing the media. The selected dimensions of the input Si waveguide are $w = 450$ nm and $h = 250$ nm as in [22] and intended for single-mode excitation. The incident wavelength is taken to be 830 nm. Another parameter we fixed is the distance between our NFT and the media which is 9 nm though this may be used as an additional parameter for fine tuning the performance. Other parameters of the structure that are optimized following the analysis outlined in the next section are listed in Table 2.

**Table 1. Refractive index, heat capacity and conductivity of the materials used in the simulation. COC stands for Carbon Overcoat**

| Parameter | Si | SiO₂ | Au | Write Pole | Air/Lube (2.5 nm) |
|---|---|---|---|---|---|
| Refractive Index (830 nm) | 3.67 | 1.45 | 0.16 + 5.15i | 2.14 + 4.24i | 1.0 |
| Heat Cap. J·m⁻³K⁻¹ | $1.63 \times 10^6$ | $1.6 \times 10^6$ | $2.49 \times 10^6$ | $3.73 \times 10^6$ | $1.0 \times 10^6$ |
| Th. Cond. W·m⁻¹K⁻¹ | 131 | 1.4 | 317 | 3.0 | 3.0 |
| **COC (5 nm)** | **Capping (1.5 nm)** | **FePt (9nm)** | **MgO (7 nm)** | **AUL (9 nm)** | **Heat sink (72 nm)** |
| 2.4 + 0.51i | 4.71 + 3.38i | 3.04 + 2.69i | 1.65 | 5.87 + 4.68i | 3.72 + 3.99i |
| $1 \times 10^6$ | $2.89 \times 10^6$ | $2.89 \times 10^6$ | $3.36 \times 10^6$ | $1.1 \times 10^6$ | $3.06 \times 10^6$ |
| 1.0 | 20 | 5.7, 1.4 (in plane) | 30 | 50 | 10.5 |

**Table 2. Initial parameters to be optimized in light-heat process**

| Parameter | $W$ | $h$ | $h_{insulator}$ | $h_{metal}$ | $w_{tip}$ | $L_{taper}$ |
|---|---|---|---|---|---|---|
| Value | 450 nm | 250 nm | To be optimized | To be optimized | To be optimized | To be optimized |

### 2.1 Optical behavior

The main parameters of the NFT that are analyzed are $w_{tip}$, $h_{metal}$, $h_{insulator}$, and $L_{taper}$ as labeled in Fig. 1. The parameter $w_{tip}$ controls the width of the taper's surface at the ABS and is therefore important to define the width of the spot within the FePt layer as will be shown below. In the same way, the thickness of the metallic taper, $h_{metal}$, determines the height of such a spot. We had found that the plasmonic mode is able to diffract upwards between the tip of the taper at the ABS surface and the media, and so, the larger $h_{metal}$ the greater spot dimension. Together, these two parameters dominate the areal bit density and strongly regulate both the optical and thermal efficiency as light couples from the NFT to recording medium. $h_{insulator}$ determines the losses of the hybrid plasmonic waveguide (for $h_{insulator} \approx 10$ nm the waveguide is in the low loss regime [37]) and is also used to match the momentum to the input Si photonic strip waveguide. Lastly, $L_{taper}$ will be used to reduce absorption losses within the Au taper and optimize the tapering angle.

To reiterate, the figure of merit in the optical simulations is the absorption efficiency defined to be the optical power within the FePt layer over a spot of $50 \times 50$ nm² divided by the input power to the Si waveguide. We use media like that previously published on HAMR devices although optimization of the media itself or write pole was not part of this study [5, 38–44]. Additionally, the distance between the write pole and heated spot within the recording layer is limited so that the applied external magnetic field effectively adjusts the magnetization in the recording layer. For these reasons, i.e. minimum spot size and pole-spot separation, the optimization process depicted in Fig. 2(a) begins with the $SiO_2$ and Au thicknesses. The latter of which is set with an upper limit of 60 nm. Furthermore, it has been suggested in [22] that matching the effective mode indexes of photonic and plasmonic modes made a dominant contribution to the efficiency for a relatively short taper of length = 500 nm ($L_{taper}$) [16, 37]. Therefore, as a first approach we select $L_{taper}$ = 500 nm and $w_{tip}$ = 50 nm before commencing the optimization process.

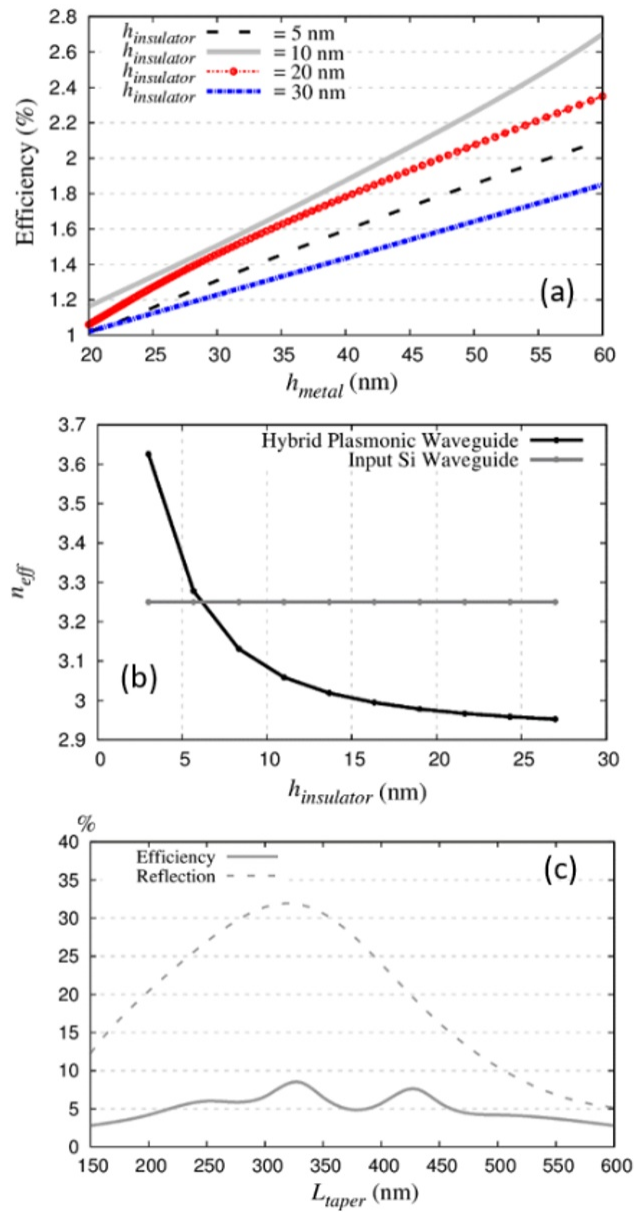

Fig. 2. (a) Efficiency versus $h_{metal}$ for optimization of the parameter $h_{insulator}$. (b) Further optimization of $h_{insulator}$ from plot of the real component of the effective mode indexes in the hybrid plasmonic waveguide and matching them to the input slab waveguide (c) optimization of the length of the NFT $L_{taper}$ for $w = 450$ nm, $h = 250$ nm, $w_{tip} = 50$ nm, $h_{insulator} = 10$ nm and $h_{metal} = 60$ nm. We show the efficiency versus the taper length with two maxima at 330 nm and 450 nm as well as the reflectivity within the Si waveguide.

A selection of our results is plotted in Fig. 2(a) where data was tested in no smaller than 5 nm increments of insulator thickness allowing for a few nanometers of surface roughness and variations in deposition technique. The best performance was obtained for $h_{metal} \approx$ 60 nm and

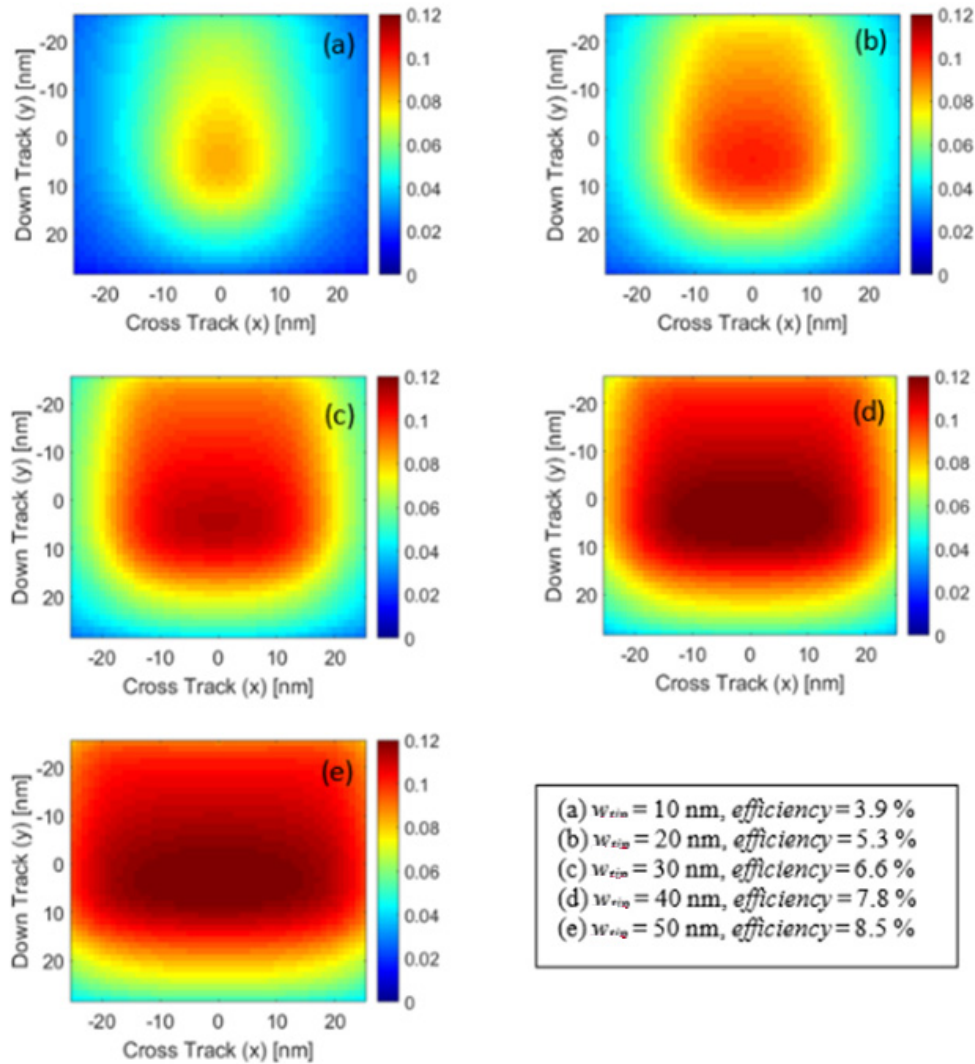

(a) $w_{tip}$ = 10 nm, $efficiency$ = 3.9 %
(b) $w_{tip}$ = 20 nm, $efficiency$ = 5.3 %
(c) $w_{tip}$ = 30 nm, $efficiency$ = 6.6 %
(d) $w_{tip}$ = 40 nm, $efficiency$ = 7.8 %
(e) $w_{tip}$ = 50 nm, $efficiency$ = 8.5 %

Fig. 3. Normalized optical power in the FePt layer for different values of wtip and the optimized values: w = 450 nm, h = 250 nm, hinsulator = 10 nm, hmetal = 60 nm, and Ltaper = 330 nm. Each legend represents the normalized optical power in the middle of the FePt layer. The horizontal axis represents the cross-track direction and the vertical axis is down track.

$h_{insulator} \approx$ 10 nm. Beyond $h_{metal} \approx$ 60 nm the optical spot becomes much larger than the desired spot size and the coupling efficiency begins to saturate since we are only considering an area of $50 \times 50$ nm$^2$ in its calculation. Although additional power may be delivered to the media outside this region for larger values of $h_{metal}$, we deem the spot too large to yield sufficient areal bit densities. With these values, the momentum of the fundamental TM mode in the input Si strip waveguide and the mode at the beginning of the hybrid plasmonic waveguide can indeed be matched [37, 45]. To check this point, we determined the real part of the effective index ($n_{eff}$) of the hybrid plasmonic waveguide versus the thickness of the insulator, $h_{insulator}$, and compared it to the effective index of the input Si waveguide with $n_{eff} = 3.25$ in Fig. 2(b). We can see that the crossing is around $h_{insulator} = 8$ nm. Even though the momentum matching is closer to $h_{insulator} = 5$ nm, the plasmonic losses for $h_{insulator} = 5$ nm are larger than for $h_{insulator} = 10$ nm and therefore the absorption efficiency is smaller. Particularly small thicknesses of the SiO$_2$ layer are sought after, where along with the relatively high refractive index of Si, it further confines the plasmon mode on the ABS surface which eventually forms the heated spot in the recording medium. By contrast, effective mode matching was used in [16] for the same conceptual design however using 1550 nm light and various Au thicknesses which required $h_{insulator} > 50$ nm and frequently over 100 nm thick. If $h_{metal} > 60$ nm, we find the spot on the surface of the media will be bigger than $50 \times 50$ nm$^2$ and efficiency begins to drop since energy is being diffracted outside the desired area for bit writing. It is also important to note that at Au thicknesses of 60 nm the plasmonic mode propagated along the bottom surface of the taper only. As thicknesses approached 40 nm, the hybrid waveguide began to deviate from this behavior and an NFT (a) symmetrical mode can be seen oscillating along both top and bottom surfaces.

Now fixing $w_{tip} = 50$ nm, $h_{metal} = 60$ nm and $h_{insulator} = 10$ nm, we scan the length of the NFT to find a maximum efficiency since it is formed by a Fabry-Perot (FP) cavity with one facet at the tip and the other facet at the opposite end of the NFT, however tapering may shift the position of FP resonances. The results are shown in Fig. 2(c), where it is possible to observe two plasmonic resonances with maxima at $L_{taper} = 330$ nm and $L_{taper} = 450$ nm. These positions are roughly separated by a half-integer of the plasmon wavelength ($\approx 270$ nm) where some deviation is expected due to the sharp tapering angle. Experimentally, oscillations related to the plasmon wavelength have been verified in long Au tapers [46]. Optimization of the taper length according to plasmon wavelength is fundamentally different than optimization via coupling length such as that seen in similar antenna-based NFTs proposed for HAMR and may extend the taper length by many microns making a less compact design [16, 46]. A maximum optical absorption efficiency around 8.5% is obtained for $L_{taper} = 330$ nm comparing well with other HAMR devices [1, 22, 25, 38, 47, 48]. Though the back reflection into the Si waveguide is nearing a maximum at this point, further alterations can be made by abutting anti-reflective coatings or anti-reflective trenches which have been shown capable of improving optical efficiencies significantly in plasmonic NFTs [49]. The interface between the input Si strip waveguide and the NFT is the source of approximately 4% of the reflection with the bulk of the remaining reflected power coming from the tip of the NFT and the media. It is anticipated that for $L_{taper} = 450$ nm there are more Ohmic losses and so the optical absorption efficiency and thermal efficiency will be reduced with respect to the tip set at $L_{taper} = 330$ nm. Therefore, to have a compact device and enhance the thermal performance in the recording layer, we select the value of $L_{taper} = 330$ nm. Consequently, the temperature reached in the material will be larger than in the case of $L_{taper} = 450$ nm for the same input power. This selection corresponds to the worst-case scenario in the thermal simulations if one is looking to reduce the temperature in a metallic taper rather than maximizing throughput to recording media. For this reason, we include the additional modification of a heat spreader to the HAMR structure as shown in Fig. 4, which will be discussed in the following subsection.

Next, the optical power spot in the FePt layer is adjusted to maximize areal bit density by modifying the value of $w_{tip}$. In Figs. 3(a)-3(e), we demonstrate how the spot can be tuned by changing $w_{tip}$ which effectively manipulates the thermal conductivity circa the FePt surface. Plotted is the optical spot within the middle of the FePt layer, which we find is a closer representation to the actual thermal spot size which is typically larger. The majority of the nanofocused light is kept within an area of $50 \times 50$ nm$^2$ for all values of the size of the taper opening. It is possible to observe that the spot is not square due to the fact the NFT is not symmetric. Although optical efficiencies are maximum for larger values of $w_{tip}$ a choice of $w_{tip}$ = 20 nm, shown in Fig. 3(b), is decided upon given the smaller spot size and higher areal density that will be yielded. Also, in general the thermal spot profile is often larger than the optical spot profile due to the cross track thermal conductivity in the media layers. Optical absorption is still high at 5.3%.

In the following section, we present the thermal study of the write head and media to determine how much input power is necessary to inject into the input Si waveguide. Since Curie temperatures may vary depending on the grain size and lattice structure of FePt, we aim to heat the media to a peak temperature above 800 K. The optimized optical parameters we use are: $w$ = 450 nm, $h$ = 250 nm, $h_{insulator}$ = 10 nm, $h_{metal}$ = 60 nm, $L_{taper}$ = 330 nm, and $w_{tip}$ = 20 nm.

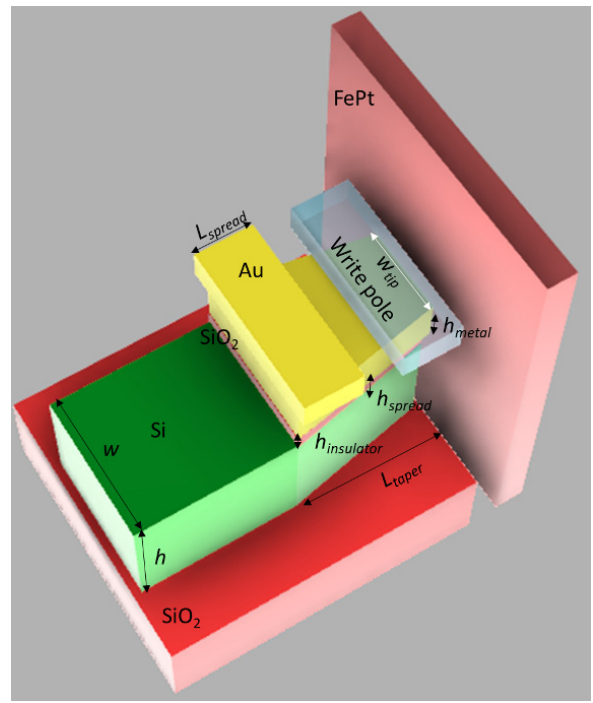

Fig. 4. Modified structure of Fig. 1 with a heat spreader in the back to reduce the temperature of the tapered metal. The dimensions of the heat spreader must be optimized to reduce the temperature of the NFT.

## 2.2 Thermal behavior

For HAMR, the writing process may be undertaken using the steady-state or pulsed excitation of the NFT. This thermal analysis uses the former, where we extract the steady-state power loss density throughout the entire structure (including cladding) from the Helmholtz equation and input this value as resistive losses within the steady-state thermal diffusion equation [38]. For this, we used COMSOL software to perform the modeling where we calculated a single mode as input to the Si waveguide and propagate it towards the NFT [50]. All simulations are in 3D. As indicated earlier, failure of the NFT is a major impediment holding back the implementation of HAMR and so we look to keep temperatures of the Au film under 400 K. Even with temperature rises of ~100 K the hardness of Au films may be reduced by half and adhesion forces contributing to film dislocation can significantly increase [51, 52]. The disk rotational speed is typically 5400 rpm and the linear speed used for the recording is roughly 20 m/s, resulting in roughly 1 ns for the heating process. As studied in [53] we performed a steady state analysis of the heating since the velocity of the disk has only a small influence on the thermal behavior which may adjust the magnetic write width by a few nanometers [53]. The disk moves in the cross-track direction.

We simulated the structure using the optimized parameters for which we found temperatures in the FePt to surpass 800 K for input powers of roughly 3.75 mW while the Au taper reached a peak temperature of roughly 430 K. To reduce the Au temperature, we modified the structure in Fig. 1 by placing a heat spreader in the NFT design as shown in Fig. 4. The spreader is a block of Au placed at the back of the tapered metal. As mentioned in the preceding subsection, only the bottom part of the taper is needed to produce the focusing while the top surface can be used to construct a heat spreader and lower the overall temperature of the NFT. This behavior allows a heat spreader to be added without incurring a negative effect on the optical absorption efficiency of the NFT.

A comparison between the design in Fig. 1 without the heat spreader and Fig. 4 with the heat spreader is shown in Fig. 5, where the temperature distribution of the device is plotted along a cross-section cutting through the center of the Au taper for an input power of 3.75 mW. The scale highlights temperatures reaching above 400 K which are colored white and temperatures below 350 K in dark red. Here it is clear that the heat spreader is very effective at maintaining the NFT temperatures below 350 K compared to the case without the heat spreader. We find the heat spreader's ability to reduce temperatures of the taper is saturated with a height ($h_{spread}$) of roughly 150 nm. The length ($L_{spread}$) is set at 222 nm leaving about 50 nm of space between the spreader and write pole in case the pole size needs to be increased. In Fig. 5(a) the maximum temperature in the Au of the NFT is 403 K with a maximum temperature of 813 K reached in the media while in Fig. 5(b) maximum temperatures reach 428 K and 803 K in the Au and media, respectively. Table 3 lists thermal efficiency parameters comparing the two cases in Fig. 5, namely, maximum temperatures in the Au and recording layer as well as thermal gradients in the FePt layer reported within a simulation error of ± 3%. Most notably is the change in temperature difference between the Au taper and recording layer between the two cases which reaches 35 K. This can yield a significant improvement in lifetime of the NFT. Values reported for the ratio of the temperature rise in the media compared to that in the Au at the steady state ($\Delta T_{\text{Media}}/\Delta T_{\text{NFT}}$) are significantly higher than those found reported for antenna-based NFTs though lower compared aperture-based designs [25, 38]. However, thermal gradients in antenna-based designs tend to be higher compared to aperture structures, for which the tapered NFT presented in this article has maximum values closely matching state-of-the-art [3, 4] values between 10 and 15 K/nm with estimations in Au thermal conductivity and thermal spot size slightly varying. We additionally simulated each case with a reduced Au thermal conductivity ($317 \text{ W/m}^{-1}\text{K}^{-1} \rightarrow 80 \text{ W/m}^{-1}\text{K}^{-1}$) in the down track direction where Au thickness is only 60 nm. It is notable that adding the heat spreader increased the maximum cross track gradients as well as slightly increased the temperature in the recording layer. This is due to a small improvement in optical efficiency (≈0.2%) which coincides with a rise of absorbed power at the backend of the Au taper plus Au heat spreader surface where the evanescent wave is incident upon.

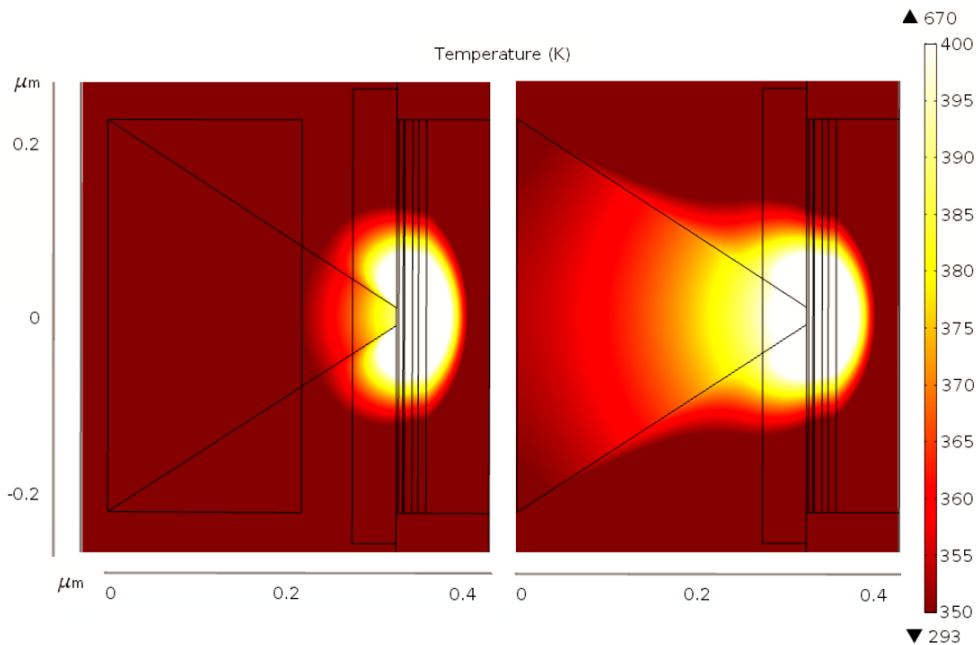

Fig. 5. Temperature distribution in the NFT taken over a cross-section that cuts through the center of the Au taper for an input power 3.75 mW with the heat spreader included (a) and without the heat spreader included (b). The same scale is used in both figures (a) and (b) demonstrating much of the Au taper is kept under 350 K using the heat spreader. Outlines of the taper, heat spreader, write pole and media are shown.

We subsequently calculate the thermal profile within the recording layer in Fig. 6(a). The full-width half-maximum (FWHM) of roughly $65 \times 70 \text{ nm}^2$ for the thermal spot is closely given by the contour at 550 K. The thermal spot is reduced to approximately $30 \times 35 \text{ nm}^2$ near the point where gradients reach a maximum between contours of 650 K and 750 K. It is at this position of elevated temperature gradients where ideally one would like to apply the external magnetic field for bit writing. The FWHM compares very well with those reported for antenna-based HAMR designs for the NFT review in [25] although no write pole had been included in their study. Noticeably, the down track temperature gradient shown in Fig. 6(b) is highly asymmetric due to the placement of the write pole leading to the elongation of the thermal spot profile in this direction. Depending on write pole design, the distance between the pole and position of the written bit is desired to be as close as possible, preferably approaching 30 nm [54]. This measure further assures consistent bit writing. From Fig. 6(b), a pole-bit distance ranging from 30 to 40 nm yields a down track gradient value of roughly 5 K/nm just above the 650 K contour. This value still compares well with those previously reported, but it should be noted that if a larger gradient is desired then the thickness of the Au can be reduced to achieve this effect. A slight reduction in optical efficiency would occur as shown in Fig. 2(a), though the sought-after improvement to the thermal gradient is accomplished. The symmetry of the cross-track gradient is maintained with maximum values reaching over 14 K/nm in our model, approximately 1.5 K/nm higher than the case shown in Table 3 without the heat spreader.

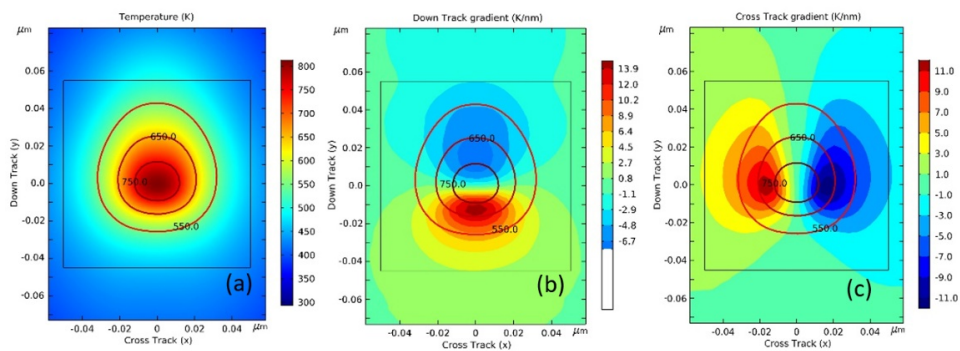

Fig. 6. (a) Thermal spot on the surface of the FePt layer for the optimized parameters of $w = 450$ nm, $h = 250$ nm, $w_{tip} = 20$ nm, $h_{metal} = 60$ nm, $h_{insulator} = 10$ nm, $L_{taper} = 330$ nm, and injecting 3.75 mW of input power. The horizontal axis represents the cross-track direction for bit writing while the vertical is down track. The write pole sits at top edge of the $100 \times 100$ nm$^2$ box shown and 9 nm into the page. Cross (b) and down track (c) gradients are shown demonstrating a desired symmetry cross track for consistent bit writing/reading. Temperature contours of 550 K, 650 K, and 750 K are shown on each plot.

**Table 3. Steady State Thermal Efficiency Parameters for the Antenna-Based Hybrid Waveguide**

| NFT Design | $\Delta T_{Media}/$ $\Delta T_{NFT}$ | Max Temp NFT | Max Temp FePt | Max Cross Track Grad. | Max Down Track Grad. |
|---|---|---|---|---|---|
| Heat Spreader | 3.8 | 403 K | 813 K | 14.1 K/nm | 15.1 K/nm |
| Heat Spreader + reduced Au Thermal Conductivity | 3.5 | 413 K | 815 K | 14.1 K/nm | 15.2 K/nm |
| No Heat Spreader | 3.0 | 428 K | 803 K | 12.5 K/nm | 15.7 K/nm |
| No Heat Spreader + reduced Au Thermal Conductivity | 2.9 | 437 K | 802 K | 12.6 K/nm | 15.7 K/nm |

## 3. Discussion

In Fig. 7(a) we plot the best fit curve demonstrating the rise in temperature of the Au taper, write pole, and FePt for a given input power. Another important thermal efficiency parameter to gauge the coupling efficiency of the device is the temperature rise of the recording layer per milliwatt, which on average is 139 K/mW for the antenna-based hybrid waveguide design. In addition, in Fig. 7(b), we report maximum thermal gradients as a function of input power and those 20 K from the peak temperature for direct comparison with the NFTs in [25]. We surpass all gradient values reported except the cross-track gradient for the E-antenna, and all values (antenna or aperture based) reported for temperature rise per milliwatt. To the best of our knowledge, these thermal data along with those presented in Table 3 exceed or closely match thermal data published for antenna-based NFTs and most data which we found reported on aperture-based NFTs. This is combined with very good optical efficiency, thus producing a highly efficient optimization process for our HAMR design. One exception was the greater temperature rise in the FePt layer reported for bow-tie apertures though as previously mentioned gradients for these structures tend to be lower. The efficiencies could be improved upon with repetition of the optimization process to help fine tune parameters while simultaneously optimizing the media. For example, increasing the capacity of the media heatsink alone was shown to more than double temperature gradients [4] and Carbon overcoating was recently reported to adjust the thermal gradient by as much as 7% [41, 42].

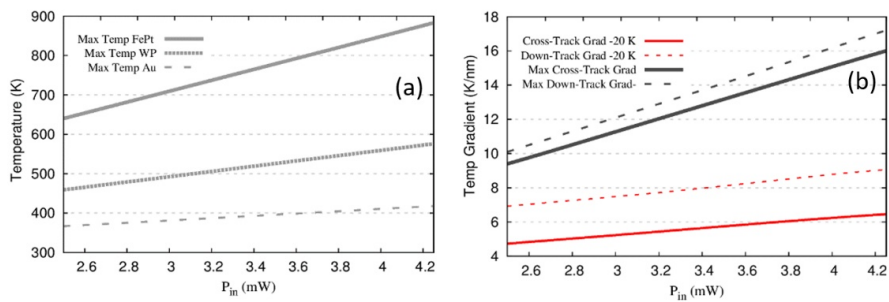

Fig. 7. (a) Maximum temperature in the FePt layer, the magnetic write pole and the NFT versus the input optical power ($P_{in}$) (b) The cross-track gradient and the down-track gradient versus the input optical power. The gray (red online) curves plot the temperature gradients at a position where temperature reaches minus 20 K from the maximum temperature in the recording layer. For comparison, values in Fig. 5 are reported for an input power of 3.75 mW.

Input power and taper length/tip width could be modified to regulate the newly adjusted temperatures. Also, depending on write pole design, NFT and write pole temperatures can be reduced further by bringing forward the heat spreader closer to the media and even abutting the pole. Moreover, the antenna-based hybrid waveguide design studied in this article only requires tapering in a single dimension and the deposition of layers no smaller than 10 nm.

There are two similar antenna-based NFTs that are used for optical nanofocusing by means of a tapered hybrid plasmonic waveguide in [16] and [22]. In the case of the structure in [16] the gold thickness of the hybrid plasmonic waveguide is 20 nm and the structure uses both the upper and bottom part of the metal, making it difficult to cool. Unfortunately, thermal simulations were not presented for these structures. The structure also has a length of 2.2 μm choosing to optimize the taper length according to the plasmonic mode's coupling length, whereas optimizing the taper length according to the plasmon wavelength enables a more compact, shorter taper to be realized. Another similar antenna-based NFT is also presented in [14]. Here the tapered Au in that structure is 30 nm thick. To achieve the efficiency they reported they needed to propagate a plasmon on the top and bottom of the tapered Au metal as in [16]. If the Au film is too thick then the top and bottom plasmons are not going to be coupled. This makes it difficult to place a heat spreader without having a significant effect on optical and thermal efficiencies. Finally, the nanobeak design, which uses a multidimensional taper in their antenna-based NFT, yields comparable optical and thermal efficiencies to the structure we present though no NFT temperatures were reported [3]. They integrate an Au block behind their Au tapered section that was reportedly optimized for optical absorption efficiency though inevitably some heat spreading will ensue. Having been integrated behind the Au taper it makes it challenging to move the Au block forward and reduce NFT temperatures without reducing the size (and possibly efficacy) of the taper. Regarding the scalability of our design, we found the plasmonic mode excited for an Au thickness of 60 nm propagates only along the bottom surface of the Au taper. This persists to a thickness of roughly 40 nm (reducing the thermal spot size) afterwhich a reduction in optical efficiency of roughly 1.5-2% percent should be expected as shown in Fig. 2(a). The Si waveguide is also able to dissipate heat through the substrate. However, the material is sligthly absorbing at 830 nm the maximum length recommended is 1.3 mm to keep efficiency reasonably high.

To conclude, we have analyzed the optical and thermal behavior of an antenna-based NFT using a tapered, hybrid plasmonic waveguide for HAMR. The optimization process succeeded in using an ultra-thin spacer layer to effectively index match the dielectric waveguide with plasmonic NFT mode. The NFT itself was then optimized to focus the plasmon wavelength using a relatively short taper of a few hundred nanometers, taking advantage of the FP-like cavity behavior allowing for a compact structure. We conclude after the thermal analysis that a heat spreader was optically and thermally beneficial for the structure under HAMR operation. Altogether, the addition of a heat spreader and the optimization process may be applied to antenna-based structures composed of other materials along with slight modifications. The optical and thermal efficiencies for the design either met or surpassed current industry standards for antenna-based NFTs with potential to be improved upon given optimization of the magnetic write pole or media [3, 4, 22, 25, 48].

### Funding

Science Foundation Ireland (SFI) (12/RC/2278, 15/IFB/3317), Western Digital Corporation, European Research Council (ERC) (713567)

### Acknowledgments

Calculations were performed at the Trinity Centre for High Performance Computing using Lumerical and COMSOL Software. We would like to acknowledge CMC Microsystems for the provision of products and services that facilitated this research, including COMSOL.